\documentstyle[prl,aps,epsfig]{revtex}

\begin{document}
\def\be{\begin{equation}}
\def\ee{\end{equation}}
\def\bea{\begin{eqnarray}}
\def\eea{\end{eqnarray}}
\def\bml{\begin{mathletters}}
\def\eml{\end{mathletters}}
\def\l{\label}
\def\eqn#1{(~\ref{eq:#1}~)}
\def\av#1{{\langle  #1 \rangle}}

\title{Locating the minimum : Approach to equilibrium in a disordered,
symmetric zero range process}

\author{Mustansir Barma and Kavita Jain}
\address{Department of Theoretical Physics, Tata
Institute of Fundamental Research,\\
Homi Bhabha Road, Mumbai 400005, India.}
\maketitle
\widetext

\begin{abstract}
We consider the dynamics of the disordered, one-dimensional, symmetric
zero range process in
which a particle from an occupied site $k$ hops to its nearest neighbour
with a quenched rate $w(k)$. These rates are chosen randomly from the
probability distribution $f(w) \sim 
(w-c)^{n}$, where $c$ is the lower cutoff. For $n > 0$, this model
is known to exhibit a phase 
transition in the steady state from a low density phase with a finite
number of particles at each site to a high density aggregate phase in
which the site with 
the lowest hopping rate supports an infinite number of particles.
In the latter case, it is interesting to ask how 
the system locates the site with globally minimum rate. We use
an argument based on local equilibrium, supported by Monte Carlo
simulations, to describe the approach to the steady state. 
We find that at large enough time, the mass transport in the regions 
with a smooth density profile 
is described by a diffusion equation with site-dependent rates, while
the isolated points where the mass distribution is singular 
act as the boundaries of these regions. Our argument implies
that the relaxation 
time scales with the system size $L$ as $L^{z}$ with $z=2+1/(n+1)$ for
$n > 1$ and suggests a different behaviour for $n < 1$.

\vskip0.5cm
\end{abstract}

\section{Introduction}
\label{intro}

The presence of quenched disorder is known to strongly affect the dynamical
and steady state properties of many systems. For instance, for noninteracting
particles moving in a random medium, disorder can lead to anomalous
transport and change the manner in which the steady state is approached
\cite{ab,kehr,ba}. Moreover, when inter-particle interactions are present,
disorder can induce new collective effects such as phase separation
\cite{krugR}. In this paper,
we study the effect of quenched, sitewise disorder on a simple
stochastic model of interacting particles known as the zero range process
\cite{spitzer,evansR}. This process can be viewed as describing a
system of interacting particles hopping in and out of wells with
various depths. While the static properties of this model are
known analytically, the temporal properties are not as well characterized.
Here we study the approach to the steady state of a
zero range process which undergoes a disorder-induced phase transition.

The study of stochastically evolving lattice models
has played a central role in better understanding 
interacting, statistical systems \cite{books}. Such models are defined
directly through simple stochastic 
rules, in contrast to the traditional route of defining the
interactions through a Hamiltonian and then constructing
dynamical rules consistent with it. The zero range process describes
one such class of lattice models; other important classes include the
asymmetric simple 
exclusion process which is a simple model of a current-carrying system
\cite{spitzer,liggett}, and the contact process whose steady state
shows a phase transition from an active phase to a dead phase where no
further evolution is possible \cite{liggett}.

The zero range process deals with a conserved number of particles
hopping on a lattice with, in general, site-dependent rates. For
almost all choices of these rates, the particles 
have on-site interactions; however, these rates do not depend on the
particle occupation at other sites. In this sense, the range of
interaction is zero. 
The steady state of this process is known to be of
product measure form in all dimensions \cite{spitzer}. In the recent 
past, the zero range process has been used to model several systems
with quenched disorder, such as 
traffic flow in a system of cars with different preferred speeds
\cite{kf} and activated flow down a rugged slope \cite{gt}, besides
several other applications \cite{evansR}. A recent application is in
a model of polymerization in a random medium with imperfect traps
\cite{dagg}. 
Interestingly, for several choices of disorder,
as the density is increased, the steady state of the system shows a
phase transition from a homogeneous phase 
in which the density is roughly uniform, to an infinite aggregate phase
in which the density profile has a singularity at the site with the
lowest hopping rate.

The latter state is particularly interesting since in this
phase, starting from an initial random distribution of particles, an
infinite aggregate is formed at the site with the lowest hopping rate.  
The question arises: what is the  
mechanism by which the system locates the site with \emph{globally
minimum} rate and transports a finite fraction of the total number
of particles to it through the random medium? In this paper, we
address this question for a symmetric zero range process in one dimension. 
Figure~\ref{tseries} shows how the density profile evolves in time
starting from an initial random distribution of particles 
(Fig.~\ref{tseries}a) to the steady state
in which an aggregate builds up at the site with the lowest hopping
rate (Fig.~\ref{tseries}d).  

The rest of the paper is organized as follows. We define the model and
discuss its steady state properties in Section~\ref{model}. In
Section~\ref{hydro}, we first give a simple, qualitative picture of the
dynamics of relaxation to the steady state. 
The scaling properties of the relaxation time define a dynamic
exponent which is determined using local equilibrium arguments,
supported by Monte Carlo 
simulations. Finally, we conclude with a discussion of open questions. 

\section{The Model and its steady state}
\label{model}

In this section, we define the model and briefly discuss its steady
state properties. We consider the unbiased or biased motion of a
conserved number $M$ of particles on a $d$ dimensional lattice of
length $L$ with periodic boundary conditions.  
At any site $k$ occupied by a nonzero number of particles, a single 
particle attempts to hop out at a rate $w(k)$, independent of the
number of particles present at site $k$ or its neighbours. Note that
this choice of rates implies an attractive on-site interaction since
the hop-out rate of these particles is lower than that of noninteracting
particles. The zero range process with this particular choice of rates
has appeared in several contexts, as in a particlewise disordered,
asymmetric exclusion process \cite{kf,bec} and as a limiting case of a
model of aggregation and fragmentation with sitewise disorder \cite{dagg}.

As discussed in Section~\ref{intro}, the problem is essentially to
understand how the system locates the site with the 
minimum hopping rate. For this reason, it is useful to assign two labels
to the hopping rate. We denote the rates by 
$w_{j}(k)$ where $k$ is the site index and $j$ is the 
index when the rates are arranged in ascending order with $j=1$
labelling the lowest rate.
Correspondingly, $m_{j}(k)$ denotes the mass (or number of
particles) at the site with
hopping rate $w_{j}(k)$ at a particular instant. In the following, we
will only display the labels which are pertinent to the discussion and
suppress the rest. Also, unless the time
dependence is explicitly specified, all the quantities will refer to the 
steady state. 

The rates $\{ w \}$ are chosen independently for all $k=1,...,L^{d}$
from a common probability distribution  
\be
f(w)=\left[ (n+1)/(1-c)^{n+1} \right] \;\;\;\; (w-c)^{n}\;,\;\;\;w \in
[c,1]\;\;\;,\;\;\;c > 0 \;\;\;,\;\;\;n > 0\;\;\;. \l{fw}
\ee
For both the symmetric and the totally asymmetric case, the
probability $P(\{ m(k) \})$ of a configuration  
$\{m(1), m(2), ...,m(L^{d})\}$ is known to be given by \cite{evansR}   
\be 
P(\{ m(k) \})=\frac{1}{\cal{N}} \prod_{k=1}^{L^{d}} \left( \frac{v}{w(k)}
\right)^{m(k)}  \;\;\;\;,  \l{soln}
\ee
where $\cal{N}$ is the normalisation constant and $v$ is the fugacity
which can be determined using the conservation law $\sum_{k=1}^{L^{d}}
m(k)=M$. This solution holds in all
dimensions and for all bias. In the presence of bias, Eq.(\ref{soln}) 
describes a nonequilibrium steady state 
whereas in the absence of bias, 
it describes an equilibrium state as the condition of
detailed balance holds. This equilibrium state is described
by a Hamiltonian which is long ranged, thus allowing the system to have a phase
transition even in one dimension \cite{evansR}. The probability $P(m,k)$ that
there are $m$ particles at site $k$ is given by 
\be
P(m,k)=\left( 1-\frac{v}{w(k)} \right) \left(
\frac{v}{w(k)} \right)^{m}  \;\;\;\;,
\ee
which implies that the average number of particles at site $k$ is 
\be
\av{m(k)}=\frac{v}{w(k)-v}=\frac{s(k)}{1-s(k)} \l{avgm} \;\;\;\;,
\ee
where $s(k) \equiv \sum_{m=1} P(m,k)=v/w(k)$ is the average occupation
probability of site $k$. Since the total number of particles is conserved, 
\be
\rho=\frac{1}{L^d} \; \frac{v}{w_{1}-v}+\int_{c}^{1} dw \;
\frac{v}{w-v} \;\;f(w) \l{cons} \;\;\;\;,
\ee
where $\rho=M/L^{d}$ is the total particle density and disorder averaging has
been done. This equation is reminiscent of 
Bose-Einstein condensation in the ideal Bose gas where the lowest
momentum state is macroscopically occupied beyond a critical
density $\rho_{c}$ \cite{bec}. 
At $\rho=\rho_{c}$, the fugacity $v$ gets
pinned to the lower cutoff of $w$, namely $c$ (to which $w_{1}$ will
tend in the thermodynamic limit) so that the site with the lowest
hopping rate supports an aggregate with 
a finite fraction of total particles. The condition $\rho-I(c)=0$
determines the critical point $\rho_{c}=c (n+1)/n (1-c)$ where $I(v)$
is the integral on the right hand side of Eq.(\ref{cons}).

For $\rho < \rho_{c}$, the typical mass at all sites 
is of order unity. As the density is increased, there is a phase transition
at $\rho=\rho_{c}$. In the high density phase, the slowest site
supports mass of $O(L^{d})$ whereas the site with rate $w_{j}$, $j
\neq 1$
supports mass of $O((L^{d}/j)^{1/n+1})$. The latter can be seen via a
simple argument. Consider a variable $x$ distributed 
uniformly between $0$ and $1$. Since on
average, $L^{d}$ observations of $x$ will be equally spaced, it follows
that the $j$th lowest observation in $L^{d}$ trials is typically at a
distance $j/L^{d}$ 
above zero. Since $\av{m_{j}}$ is inversely proportional to the
separation between the lowest and the $j$th lowest hopping rate and 
$f(w)$ can be related to the uniform distribution by a change of variables
($w=c+(1-c) \;x^{1/n+1}$), one obtains $\av{m_{j}} \sim (L^{d}/j)^{1/n+1}$,
$j \neq 1$.

\section{Approach to the steady state}
\label{hydro}

In this section, we will discuss the approach to the steady state for the
symmetric, zero range process in one dimension with the hopping rates
chosen from $f(w)$ defined in Eq.(\ref{fw}). We first describe a
simple picture of approach to the steady state. Then we present an
argument based on local equilibrium which
suggests that the relaxation time scales with system size $L$ 
as $L^{z}$ with $z=2+1/(n+1)$ for $n > 1$. However, this treatment 
breaks down for $n < 1$.

\begin{figure}
\begin{center}
\psfig{figure=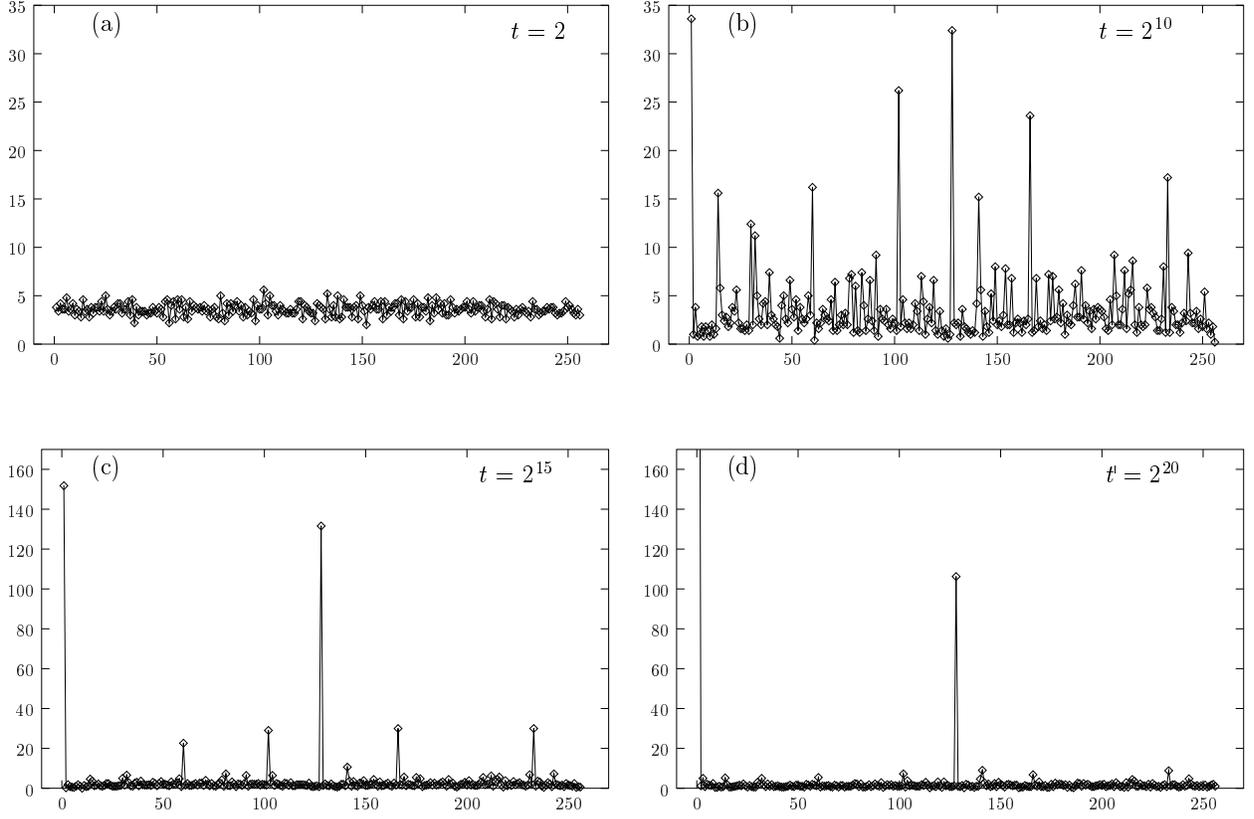,width=17cm,angle=0}
\caption{Density profile $\av{m(k,t)}$ vs. $k$ in the aggregate phase
for a given realization of disorder at time 
$(a) \; t=2, (b)\; t=2^{10}, (c) \;t=2^{15}, (d) \; t=2^{20}$. The
site with the lowest hopping rate 
is located at $k=1$ and the second lowest at $k=L/2$. At $t=2^{20}$,
the system is close to the steady state and
$\av{m_{1}(1,t)}=426$. Parameters used: $L=256, \rho=4, n=2, c=0.5.$}  
\label{tseries}
\end{center}
\end{figure}

\begin{figure}
\begin{center}
\psfig{figure=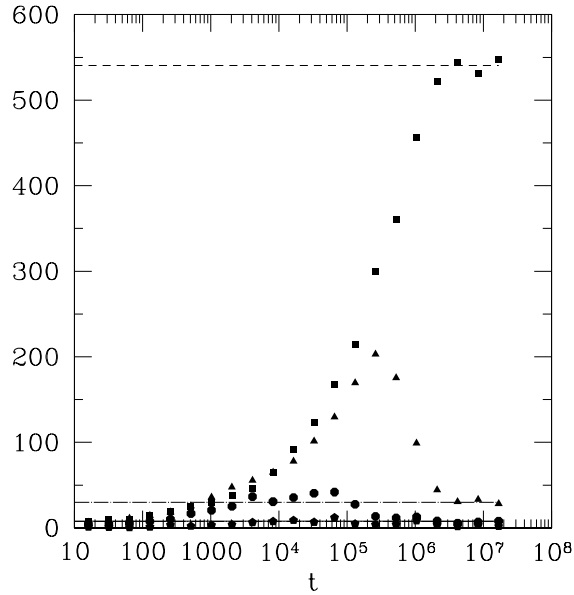,width=8cm,angle=0}
\vspace{0.1in}
\caption{Time dependence of the average density $\av{m_{j}(t)}$ for
four slowest sites namely $j=1$ (squares), $2$ (triangles), $3$
(circles) and $4$ (pentagons) for a given disorder configuration. The
horizontal lines denote the steady state value for $\av{m_{j}(t)}$ calculated
using Eqs.(\ref{avgm}) and (\ref{cons}). Parameters used:
$L=256, \rho=4, n=2, c=0.5.$}  
\label{4low}
\end{center}
\end{figure}

We first illustrate qualitatively the temporal sequence of events
through which an aggregate with mass of $O(L)$ is formed at the site with
the lowest hopping rate, starting from a random initial
condition in which each site has mass of order unity
(Fig.~\ref{tseries}a). The behaviour of the average mass 
$\av{m_{j}(t)}$ for $j=1,...,4$ as a function of time is shown in
Fig.~\ref{4low}. Here $\av{...}$ 
is to be understood as an ensemble average over evolution histories.
We find that
$\av{m_{1}(t)}$ rises steadily and then saturates to its steady state
value while each of $\av{m_{2}(t)}$, $\av{m_{3}(t)}$ and $\av{m_{4}(t)}$ 
rise, attain a maximum and then decay to their respective
steady state values. This non-monotonic behavior is not hard to
understand. 
At some finite time, each particle is able to move only a finite
distance away from its initial position and tends to get temporarily 
trapped at the site with the lowest $w$
within the neighbourhood explored by it, rather than the
global minimum $w_{1}$. A typical configuration at such intermediate
time thus has several large aggregates at such local minima while the 
rest of the system has masses of order unity
(Fig.~\ref{tseries}b). This explains why  
at short enough times, $\av{m_{j}(t)}$ for $j=1,...,4$ are of the same
order and increasing (Fig.~\ref{4low}). As time progresses, the
particles are able to access larger regions in 
space and identify new local minima. Then the mass increases at these
newly accessed local minima at the expense of the previous ones
(Fig.~\ref{tseries}c). This explains the drop in  
$\av{m_{4}(t)}$ and $\av{m_{3}(t)}$ after they have reached their
respective peak values, though $\av{m_{1}(t)}$ and $\av{m_{2}(t)}$
continue to rise (Fig.~\ref{4low}).
Finally, $\av{m_{2}(t)}$ also starts dropping and the excess mass is
transported to the location of the global minimum 
(Fig.~\ref{tseries}d). Once the global minimum is 
recognised by all the particles, the system reaches a steady state and
$\av{m_{j}(t)} \rightarrow \av{m_{j}}$ for all $j$. 

We now turn to an analytical description of the mechanism
of the mass transport. The exact time evolution equation obeyed by
$\av{m(k,t)}$ for a given $\{ w\}$ can be written as
\be
\frac{\partial{\av{m(k,t)}}}{\partial t}= w(k-1) s(k-1,t) + w(k+1)
s(k+1,t) -2 w(k) s(k,t) \;\;\;\;.  \l{rho}
\ee
At large enough times, the system is expected to be in \emph{local}
equilibrium.  
This allows one to assume the $s(k,t)-\av{m(k,t)}$ relation to be
approximately as in the steady state (Eq.(\ref{avgm})). Substituting
for $s(k,t)$ in the above equation, 
one obtains  
\be
\frac{\partial{\av{m(k,t)}}}{\partial t}=G(k-1,t)+G(k+1,t)-2 G(k,t)
\;\;\;\;, \l{current}
\ee
where $G(k,t)=w(k) \av{m(k,t)}/(1+\av{m(k,t)})$. 

The treatment so far is valid for any density, but we are primarily interested
in densities for which the system is in the infinite aggregate
phase in the steady state.
For such densities, at large times, one needs to divide the system into two
sets -- set $A$ composed of those sites which support
rather large aggregates and set $B$ at which the masses are small
and close to their steady state values (see Fig.~\ref{4low}). The
elements of these sets are 
not fixed in time; with the passage of time, the number of elements in
set $A$ reduce and that of set $B$ increase by the mechanism described
at the beginning of this section. Eventually, when the system is close
to the steady state, set $A$ is left with a few isolated sites
that support aggregates whose masses scale as a nonzero power of $L$ while
the bulk of the sites belong to set $B$.

We are interested in the fluctuations in the density profile about the steady
state, $\Delta m (k,t)=\av{m(k,t)}-\av{m(k)}$. 
Since $\Delta m$ is small for $k \in B$ and very large for $k \in
A$, an expansion in $\Delta m (k,t)$ would be justified only for the
background sites.
Below we first analyse these background ($B$) sites and then treat the
isolated sites in set $A$ as the boundaries of the $B$ regions.
For the $B$ sites, we may retain the lowest order terms in an
expansion of Eq.(\ref{current}) in powers of $\Delta m$ and obtain
\be
\frac{\partial{\Delta m(k,t)}}{\partial t}=D(k-1) \; \Delta
m(k-1,t) + D(k+1) \; \Delta m(k+1,t) -2 D(k) \; \Delta m(k,t)
\;\;\;\;,\;\;\mbox{$k \in B$} \;\;\;\;,\l{deltar} 
\ee
with 
\be
D(k)=(w(k)-c)^2/w(k) \;\;\;\;. \l{dx}
\ee
Here we have used the fact that $v$ tends to $c$ in the infinite
aggregate phase. 

Equation (\ref{deltar}) describes a random walker in a random
medium with site-dependent hopping rate $D(k)$. Using Eqs.(\ref{fw})
and (\ref{dx}), we find that these rates are distributed according to 
the probability distribution $g(D) \sim D^{(n-1)/2}$ for $D \rightarrow 0$.
Note that whereas $f(w)$ has a nonzero lower cutoff, $g(D)$ has zero
as the lower limit due to which it diverges as $D \rightarrow 0$
for $n < 1$. 
The problem of a particle moving with random, spatially inhomogeneous
hopping rates is well studied and has been reviewed in
\cite{ab}. For a configuration of randomly distributed $\{ D \}$ in a
large system observed on large enough time scales, 
the mean squared displacement of the random walker grows as ${\cal{D}}
t$ where the diffusion constant ${\cal{D}}=1/\av{\av{1/D(k)}}$ and 
$\av{\av{...}}$ stands for disorder average, provided $\av{\av{1/D(k)}}$ is
finite \cite{kehr}. Thus for large spatial and time separations, the particle 
diffusion is described by a single, effective diffusion constant.
In our case, one can calculate $\av{\av{1/D}}$ by noting that 
$\av{\av{1/D}}=\partial{I}/\partial{v}|_{v \rightarrow c}$. Expanding
$I(v)$ for $v$ close to $c$, we obtain  
\bea
I(v) &=& \rho_{c} + O(c-v) \;\;\;\;\;\;\;,\;\;\;\;n > 1 \\
     &=& \rho_{c} + O((c-v)^{n}) \;\;\;,\;\;\;\;0 < n < 1  \;\;\;\;,	
\eea
which implies that $\av{\av{1/D}}$ is finite for $n > 1$ and diverges for
$n < 1$. The divergence in the 
latter case indicates anomalous diffusion i.e. the mean squared
displacement grows sublinearly. In the remainder, we will restrict
ourselves to $n > 1$ \cite{comment2}.

On averaging over disorder configurations, for a system of size $L$, the
mass $\av{\av{m_{1}(t,L)}}$ on the site with lowest hopping rate is
expected to scale as 
\be
\av{\av{m_{1}(t,L)}} \approx t^{\beta} H\left(\frac{t}{L^{z}}\right)
\ee
where 
\be
H(x) \sim \cases{ \mbox{constant} & {for $x \ll 1$} \cr
	    	x^{-\beta} & {for $x \gg 1$}
}  \;\;\;\;.
\ee
Since $\av{\av{m_{1}(t,L)}} \sim L$ in the steady state, it follows that
$\beta z=1$. We now determine the relaxation time of the system by 
estimating the time $T \sim L^{z}$ required for $\av{m_{1}(t,L)}$
to reach its steady state value for a typical realisation of disorder. At
large enough times, the background sites in set $B$ form the bulk of
the system whereas only a few isolated sites with low hopping rates
belong to set $A$. Let us now consider time scales above which the two
sites with the lowest rates (i.e. $w_{1}$ and $w_{2}$) are the only
elements left in set $A$. As shown in Fig.~\ref{4low}, 
$\av{m_{2}(t)}$ drops after it has reached its peak while
$\av{m_{1}(t)}$ keeps rising at the expense of the former so that there 
is a net transfer of mass from the site with $j=2$ to the site with $j=1$.
One can think of these two sites as the boundaries to the bulk
background region with 
the former site feeding particles at rate $w_{2}$ to the bulk and 
the latter at rate $w_1$. 
On large time scales, a quasi-equilibrium is established, and in
view of the discussion above, the particles
diffuse through the bulk with an effective diffusion constant
${\cal{D}}$. Thus there is a transfer of mass at a rate ${\cal{D}} \Delta
w_{12}/r_{12}$ where $\Delta w_{12}=w_{2}-w_{1}$ and $r_{12}$ is 
the spatial separation between the two sites. The relaxation
time $T_{12}$, which is essentially the time taken to transfer the
peak mass $m_2$ at the site with rate $w_{2}$, is given by 
\be
T=T_{12}=\frac{r_{12} \; m_{2}}{{\cal{D}} \;\Delta w_{12}} \l{t12} \;\;\;\;.
\ee

Let us estimate the size dependence of the quantities that appear on
the right hand side of the above equation.
Typically the separation $r_{12}$ is of order $L$ in which case
$m_{2}$ is also of order $L$. This is because for times 
earlier than when $m_{2}$ peaks, the site $j=2$ is the slowest one
encountered by the particles in a finite fraction of the
system. Consequently, the numerator in the above equation is
proportional to $L^{2}$. This result holds even in the exceptional
case when the sites with $j=1$ and $j=2$ are separated by a distance
of order unity, as they behave effectively as a single slow site and
one can use the same reasoning as above on replacing the site with
$j=2$ by that with $j=3$.

\begin{figure}
\begin{center}
\psfig{figure=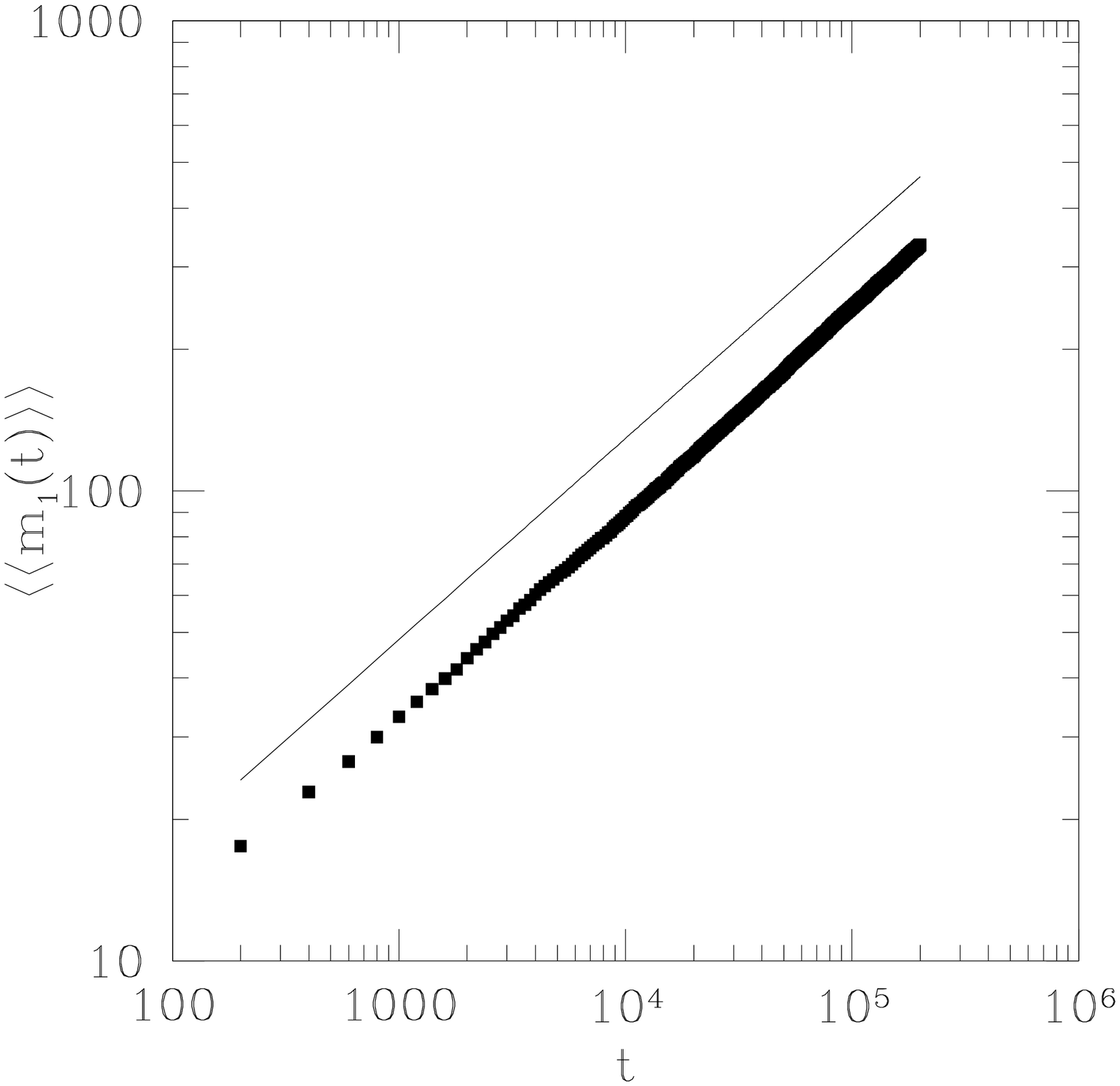,width=10cm,angle=0}
\caption{log-log plot of $\av{\av{m_{1}(t,L)}}$ vs. $t$ to show the growth
exponent $\beta$. The data for $\av{\av{m_{1}(t,L)}}$ has been averaged
over $50$ histories and $21$ disorder configurations. The theoretical
prediction for $\beta=0.428$ for $n=2$ is plotted as a solid line for
comparison. Parameters used: $L=16384, \rho=4, c=0.5.$} 
\label{beta}
\end{center}
\end{figure}

Further, using the argument given at the end of Section~\ref{model}, we find
that the inverse rate separation $(\Delta w_{12})^{-1}$ scales as
$L^{1/n+1}$. Collecting all the dependences, it follows
that the relaxation time $T$ scales with system size $L$ as $T
\sim L^{z}$ where
\be
z=2+1/(n+1)\;\;\;\;,\;\;\;\;n > 1\;\;\;\;. \l{z}
\ee
We measured the growth of $\av{\av{m_{1}(t,L)}}$ using Monte Carlo
simulations and find that it grows as $t^{\beta}$. 
As shown in Fig.~\ref{beta}, our expression for $\beta=1/z=(n+1)/(2 n+3)$ is 
consistent with the numerics. 

We conclude this paper with a discussion of two open questions: $(i)$ 
We have seen that $\av{\av{1/D}}$ diverges for $n < 1$ indicating
anomalous diffusion, which suggests that the expression for $z$ in
Eq.(\ref{z}) may be invalid for $n < 1$. It would be interesting to
find how the system relaxes in this case. 
$(ii)$ For the case with asymmetric hopping rates,
using the arguments analogous to those given above, we would expect
the dynamic exponent to be $z=1+1/(n+1)$ for $n > 1$. Once again, we may
anticipate a different result for $n < 1$. 
In \cite{kf} and \cite{kt} the dynamics of the asymmetric version has
been analysed using numerical simulations and extremal
statistics arguments for a related deterministic model. The expression
for the dynamic exponent in \cite{kf,kt} is the same as quoted above but
their arguments do not make a distinction between the regimes below
and above $n=1$. 
The elucidation of the anomalous regime for both the symmetric and
asymmetric zero range process remains an interesting open problem.

We will not attempt a summary of this paper. Professor N. Kumar once
pointed out that since it is a part of the whole, a proper summary
should include a summary of the summary and a summary of that summary,
and so on. We would rather not try ! We are very pleased that this
paper will appear in the Festschrift volume for Prof. Kumar, whose own
work has brought out many of the surprises that disordered systems
have to offer.

\begin{thebibliography}{999}
\bibitem{comment2}
Note that Eqs.(\ref{deltar}) and (\ref{dx}) are valid for all $k$ and
$n$ in the low density phase since $\av{\av{1/D}}$ always remains
finite due to $v$ being strictly less than $c$ in this phase.
\end{thebibliography}
\end{document}